\documentclass[fleqn,10pt]{wlscirep}
\usepackage[utf8]{inputenc}
\usepackage[T1]{fontenc}
\usepackage{bm}
\usepackage{hyperref}
\usepackage{svg}
\usepackage{color}
\usepackage{gensymb}
\usepackage{nicefrac}

\title{Beyond multi-view deconvolution for inherently-aligned fluorescence tomography}

\author[1,*]{Daniele Ancora}
\author[1,2]{Gianluca Valentini}
\author[1,2]{Antonio Pifferi}
\author[1,2]{Andrea Bassi}
\affil[1]{Dipartimento di Fisica, Politecnico di Milano, Piazza Leonardo da Vinci 32, 20133 Milano, Italy}
\affil[2]{Istituto di Fotonica e Nanotecnologie, Consiglio Nazionale delle ricerche, Piazza Leonardo da Vinci 32, 20133 Milano, Italy}
\affil[*]{corresponding author: \href{mailto:daniele.ancora@polimi.it}{daniele.ancora@polimi.it}}

\begin{abstract}
In multi-view fluorescence microscopy, each angular acquisition needs to be aligned with care to obtain an optimal volumetric reconstruction.
Here, instead, we propose a neat protocol based on auto-correlation inversion, that leads directly to the formation of inherently aligned tomographies. 
Our method generates sharp reconstructions, with the same accuracy reachable after sub-pixel alignment but with improved point-spread-function.
The procedure can be performed simultaneously with deconvolution further increasing the reconstruction resolution.
\end{abstract}

\begin{document}

\flushbottom
\maketitle

\thispagestyle{empty}

\section*{Introduction}
The field of Computed Tomography (CT) experienced a silent revolution during the last decade. 
A strong demand driven by deep-learning and data-mining has prompted hardware manufacturers to improve computing performances while keeping the price affordable. 
Nowadays, high throughput computation is possible with graphic-card units (GPU).
GPUs allow parallel data-processing with performances beyond belief just a few years ago, radically changing the field of signal processing. 
In particular, standard image processing tasks such as Fourier-transformation, convolution, and matrix operations experience a constant-rate performance increase \cite{sun2019summarizing,leiserson2020there}. 
GPUs are the ideal solution for the massive image processing tasks required by CT \cite{despres2017review}.
At visible wavelengths, optical projection tomography (OPT) is an example of a CT technique applied for tomographic studies at microscopic level \cite{sharpe2002optical}. 
By rotating the specimen and collecting its optical projections at multiple angles, it is possible to form the reconstruction of the specimen via CT inversion. 
Another optical technique, light-sheet fluorescence microscopy (LSFM), offers a straightforward way to optically section the sample, for the inspection of its internal structure \cite{verveer2007high}.
Even if LSFM is a direct tomographic technique (i.e. it does not strictly require computation to generate a section of the sample) it is often desirable to observe the object from different angles to increase the quality of the reconstruction. 
LSFM suffers from non-isotropic resolution (the axial resolution is lower that the lateral) and in many cases, the sample is not visible as a whole, due to tissue scattering or absorption.
Multi-view approaches address these problems, either relying on the sample rotation \cite{krzic2012multiview} or exploiting multiple objectives to observe the specimen from different angles \cite{weber2012omnidirectional}. 
Before their fusion, each acquisition is registered (aligned) against a chosen reference \cite{swoger2007multi}, to correctly overlap the information captured at different angles.
The registration is usually accomplished by locating the best overlap between the volumes, eventually including beads around the specimen to enforce the alignment fidelity \cite{preibisch2014efficient}.
Here we discuss a new reconstruction strategy for the formation of an inherently-aligned tomographic view of a specimen. 
We exploit the property of the auto-correlation (we indicate it by the operator $\mathcal{A}$) to avoid any alignment procedure. 
At the same time, we demonstrate that the auto-correlation based reconstruction brings an improved resolution, due
to the rejections of second order-correlations of the point-spread-function in the $\mathcal{A}$-space.
The work is inspired by previous results in OPT, where the auto-correlation is used to perform alignment-free reconstructions \cite{ancora2017phase}, because $\mathcal{A}$ commutes with the projection operator \cite{ancora2018optical}. 
Here, instead, we calculate a tomographic auto-correlation of the sample based on multi-view light-sheet acquisitions.
Fusing them, leaves us with an ensembled $\mathcal{A}$, created without aligning the views. 
This constitutes our starting point for the reconstruction: by inverting $\mathcal{A}$, we form a tomographic view aligned at sub-pixel level.
Furthermore, we prove that this inversion turns into a reconstruction which is sharper than the average fusion carried out in direct space. 
Since it is desirable to take into account the resolution-loss determined by the finite aperture of the optical system, our protocol can further accomplish simultaneous deconvolution with a modified Bayesian $\mathcal{A}$ inversion. 
To implement our protocols, the use of powerful GPUs is essential, otherwise this problem would just remain a mere theoretical exercise.
\begin{figure}[t!]
\centering
\includegraphics[trim=80 0 80 50, width=5.8in]{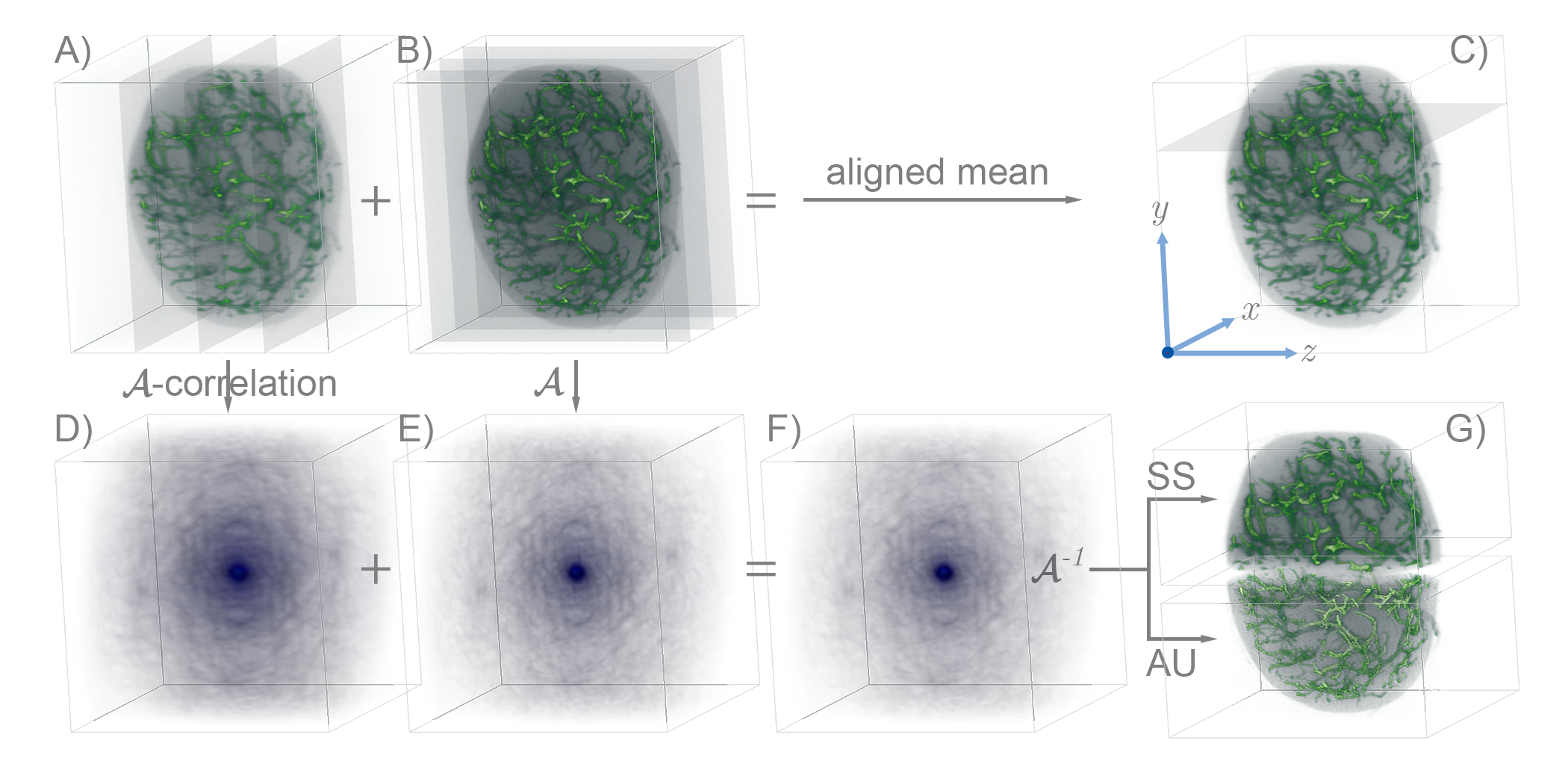}
\caption{Reconstruction pipeline.
A) Rendering of the reference view taken at $\varphi = 0\degree$. The planes indicate the $xy$-camera acquisition along the $z$-scanning direction.
B) Orthogonal detection by rotating the sample at $\varphi = 90\degree$.
C) Aligned-average of $12$ measurements. The axes are taken according to the reference view ($x$-lateral, $y$-transverse, $z$-longitudinal)
D) Auto-correlation of the view at $0\degree$.
E) $\mathcal{A}$ of the view at $90\degree$.
F) $\mathcal{A}$ averaged through $12$ angles.
G) Reconstructions obtained by using deauto-correlation methods
For visual comparison, the upper part shows the result using the Schultz-Snyder protocol, the bottom one compares it with that of the Anchor-Update method.
}
\label{fig:fig1schematics}
\end{figure}

\section*{Results}
Our reconstruction strategy is grounded on the property of the auto-correlation, that is centered in the shift-space. 
Each observation of the object is auto-correlated and concurs to the formation of the tomographic average $\mathcal{A}$ of the sample. 
Let us use the subscript $\mu$ to indicate the stack obtained by camera acquisitions, and the superscript $\varphi_i$ to denote its angular orientation indexed by $i$. 
In an experimental measurement, we observe a blurred version of the object due to the point spread function (PSF) of the system $h$, further corrupted by the presence of the noise $\varepsilon$. 
We assume an additive $\varepsilon$ that can be neglected in case of high signal-to-noise ratio measurements.
Now, let us arrange the auto-correlation in a more convenient form.
Applying the operator $\mathcal{A}$ to a given stack (\textcolor{blue}{Supplement Materials}), we have that:
\begin{align}
    \chi_\mu &\equiv \mathcal{A}\{o_\mu\} = o_\mu \star o_\mu = \left(o*h \right) \star \left(o*h \right) \label{eq:blurredACorr1}\\
    &= \chi * \mathcal{H} = o * \mathcal{K} \label{eq:blurredAcorr3}.
\end{align}
Here, $\chi=o\star o$ is the ideal auto-correlation of the object, $\mathcal{H}=\mathcal{A}\{h\}$ is the PSF in auto-correlation space and $\mathcal{K} = o \star \mathcal{H}$.
The first equality in Eq. \ref{eq:blurredAcorr3} implies that the auto-correlation of the ideal object is blurred by a kernel $\mathcal{H}$ given by the auto-correlation of the direct space PSF $h$.
The second indicates that $\chi_\mu$ can be seen as a convolution of the object with a blurring kernel that contains itself.
We consider $N$ evenly rotated measurements that we denote with the index $\varphi_i$.
The only pre-processing step required is the rotation of each measurement back to the reference angle $0\degree$ by $-\varphi_i$. 
We subtract the mean value of a dark region where the sample is not present.
Denoting each observation as $o^\varphi_\mu$, and its corresponding $\mathcal{A}$ as $\chi^\varphi_\mu$, the quantities of interest
are the averages:
\begin{equation}
   \overline{o}_\mu = \frac{1}{N} \sum_{i=0}^{N} o_\mu^{\varphi_i}, \quad \overline{\chi}_\mu = \frac{1}{N} \sum_{i=0}^{N} \chi_\mu^{\varphi_i} \quad \text{and} \quad \overline{\mathcal{H}} = \frac{1}{N} \sum_{i=0}^{N} \mathcal{H}^{\varphi_i}.
\end{equation}
Before computing the fusion as $\overline{o}_\mu$ (we use this as the ground truth), each measurement requires a perfect alignment against the reference.
The $\overline{\chi}_\mu$ is accurate, instead, because auto-correlations are centered by definition.
Ideally, this implies that we can obtain an intrinsically aligned average-reconstruction from $\overline{\chi}_\mu$, provided that we have a robust way to carry out the inversion $\overline{o}_\rho = \mathcal{A}^{-1}\{\overline{\chi}_\mu\}$.
This problem falls within the class of phase retrieval (PR), since we have access to the Fourier modulus of a real object but the phase information is missing \cite{shechtman2015phase}. 
\begin{figure}[b!]
\centering
\includegraphics[trim=0 250 0 30, width=5.in]{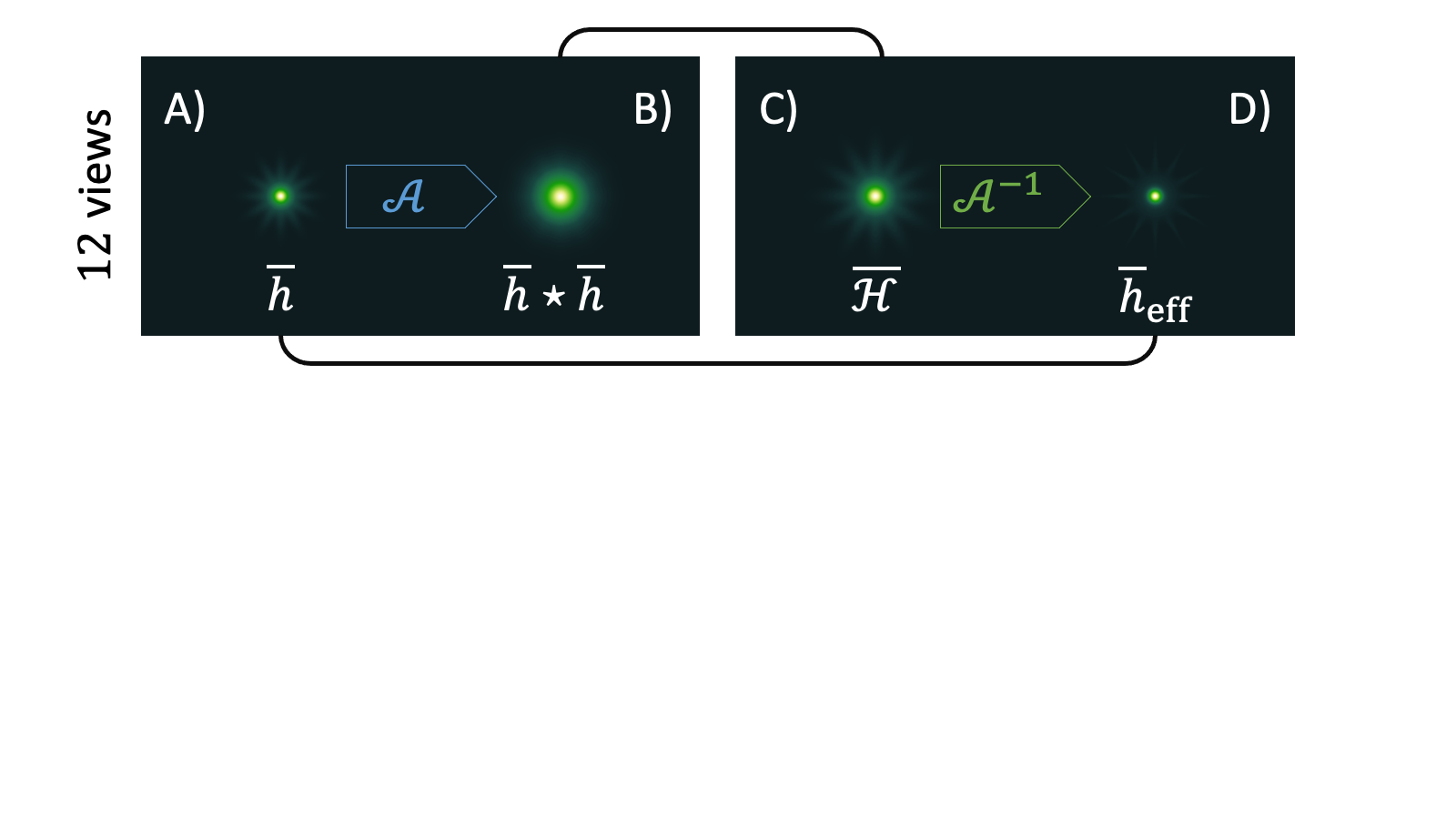}
\caption{Point-spread-functions analysis. 
A) PSF that blurs $\overline{o}_\mu$.
B) Auto-correlation of $\overline{h}$.
C) PSF $\overline{\mathcal{H}}$ that blurs the average $\overline{\chi}_\mu$.
D) Corresponding PSF in the object domain, sharper than $\overline{h}$.
}
\label{fig:fig2psfanalysis}
\end{figure}
\begin{figure*}[t!]
\centering
\includegraphics[width=5.8in]{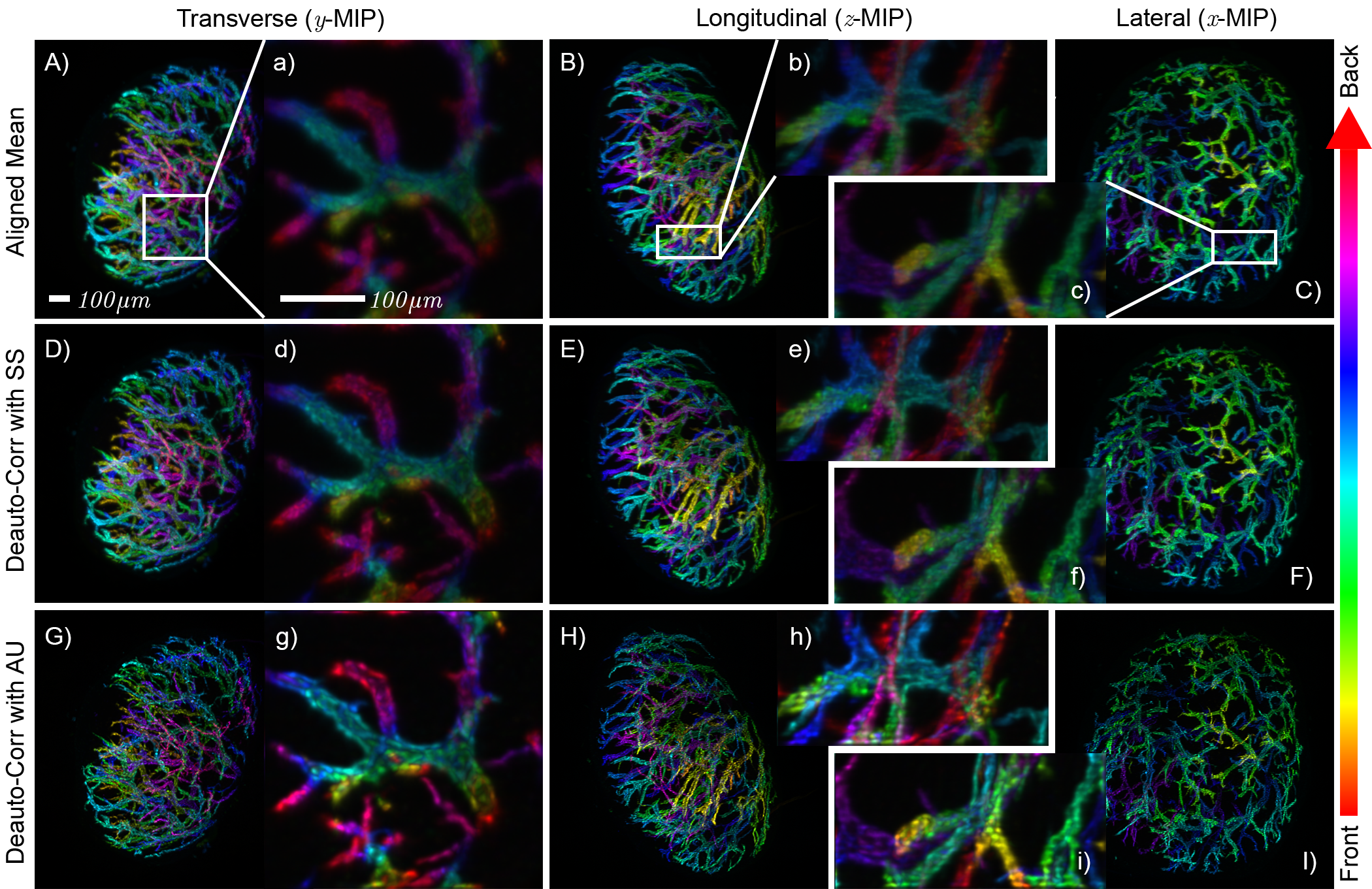}
\caption{
Comparison of the methods used to reconstruct a mouse popliteal lymph node vasculature. 
The quality of the first row improves in the central and in the bottom rows.
Results of aligned mean shown in a transverse (A), longitudinal (B) and lateral (C) direction.
The shown data are color-encoded maximum intensity projection (MIP) \cite{schindelin2012fiji} along each spatial coordinate. 
In all these MIPs, the color indicates the depth at which the
corresponding feature is located. 
The small letters indicate the cropped volume .
The cropped regions located within the whole specimen are framed with white boxes.
The scale bar is $100 \mu m$.
A,a) Transverse view of the volume $\overline{o}_\mu$ averaged and aligned by cross-correlation (Rendered in Fig.\ref{fig:fig1schematics}, viewed from the top).
B,b) Longitudinal (or side) view projection.
C,c) Lateral (or front) view.
D,d) Transverse, E,e) longitudinal, and F,f) lateral projections of the volume $\overline{o}_\rho$ reconstructed with SS.
G,g) Transverse, H,h) longitudinal, and I,i) lateral projections of the volume $\overline{o}$ deconvolved with AU.
}
\label{fig:fig2maxprojection}
\end{figure*}
With these two quantities in hand, we try to extrapolate intrinsically-aligned reconstructions by inverting the $\mathcal{A}$ with two schemes:
\begin{enumerate}
  \item[I).] Given $\overline{\chi}_\mu = \overline{o}_\rho \star \overline{o}_\rho$, find $\overline{o}_\rho$; \label{item:scheme(a)}
  \item[II).] Given $\overline{\chi}_\mu = \left(\overline{o} \star \overline{o} \right) * \overline{\mathcal{H}}$, deblur it by $\overline{\mathcal{H}}$ and find $\overline{o}$. \label{item:scheme(b)}
\end{enumerate}
At a first glance, only the second scheme that deconvolves the PSF appears to provide a super-resolved reconstruction.
However, scheme I) implies something even more interesting that we describe in Fig. \ref{fig:fig2psfanalysis}. 
By averaging $o_\mu^{\varphi_i}$ in direct space, the resulting volume gets blurred by an average PSF given by $\overline{h} = \nicefrac{1}{N} \sum_i h^{\varphi_i}$ (Fig. \ref{fig:fig2psfanalysis}A).
Its $\mathcal{A}\{\overline{h}\}=\overline{h}\star \overline{h}$ can be visualized in Fig. \ref{fig:fig2psfanalysis}B.
By averaging auto-correlations, instead, we neglect second order cross-terms of the PSF which introduce long-correlations in the fused image.
For comparison, the corresponding $\mathcal{A}$-PSF is shown in Fig. \ref{fig:fig2psfanalysis}C.
As a consequence, by solving for $\overline{o}_\rho$, we achieve an effective PSF that is sharper than $\overline{h}$.
Interestingly, this is an implicit property that comes along with the average of multiple views of $\mathcal{A}$, thus the knowledge of the PSF is not required at all.
However, for comparison, we show the effective point-spread function achieved $\overline{h}_\text{eff}=\mathcal{A}^{-1}\{\overline{\mathcal{H}}\}$ in Fig. \ref{fig:fig2psfanalysis}D.
For a detailed discussion, see the  \textcolor{blue}{Supplement Materials}.
We decided to tackle the scheme I) by using the Schultz-Snyder (SS) iterations \cite{schulz1992image}:
\begin{equation}
    o^{t+1} = o^{t} \left[ \left(\frac{\chi_\mu}{o^{t} \star o^{t}}\right)*\widetilde{o}^{t} + \left(\frac{\chi_\mu}{o^{t} \star o^{t}}\right) \star o^{t} \right]. \label{eq:SS}
\end{equation}
For the scheme II), instead, we implement the Anchor-Update (AU) protocol \cite{ancora2020IEEE} that was developed ad-hoc for this purpose:
\begin{equation}
    o^{t+1} = o^{t} \left[ \left(\frac{\chi_\mu}{o^{t} * \mathcal{K}^{t}}\right)*\widetilde{\mathcal{K}}^{t}\right], \quad \text{ updating: }
    \mathcal{K}^{t} = o^t \star \mathcal{H}. \label{eq:anchorUpdate2}
\end{equation}
Both are fixed-point iterative Bayesian methods, having the number of iterations as the only parameter to set.
In the present case, we set a high number of $5 \cdot 10^5$ iterations for both, since these methods are extremely stable and can withstand long runs.
This is also a drawback, since these algorithms suffer from slow convergence-rate because each update $t+1$ is close to the previous one $t$.

To delve into the proposed method, let us consider the volumetric acquisition taken with an LSFM setup of a cleared mouse popliteal lymph node. 
We are interested in reconstructing the three-dimensional vasculature, which was stained with a fluorescent label.
The stack $o_\mu^\varphi$ constitutes a single volumetric view of the specimen and contains the camera detections of the sample scanned through the light sheet.
We use $12$ volumes by rotating the sample in steps of $30\degree$.
In Fig. \ref{fig:fig1schematics}A-B, we render $o_\mu^\varphi$ acquired at $0\degree$ and $90\degree$.
In a normal multi-view reconstruction algorithm, every dataset has to be aligned against the reference view (we assume this to be at $\varphi = 0\degree$).
A consolidated strategy (accurate at pixel level) is to locate the maximum of the cross-correlation between the reference and the view, translating it back accordingly.
However, the researcher may be looking for sub-pixel accuracy, which would require to upsample the volume accordingly to the resolution that he wants to reach \cite{guizar2008efficient}.
This makes the size of the problem rapidly explode, leaving the user with an up-sampled estimation (with respect the original measurement) that needs to be down-sampled for the formation of the final image.
Here instead, we produce a multi-view reconstruction that is accurate at sub-pixel level and directly formed at the original resolution.
We do not calculate any volume translation, we simply process the reconstruction altogether starting from its auto-correlation $\overline{\chi}_\mu$.

\begin{figure}[t!]
\centering
\includegraphics[width=5.in]{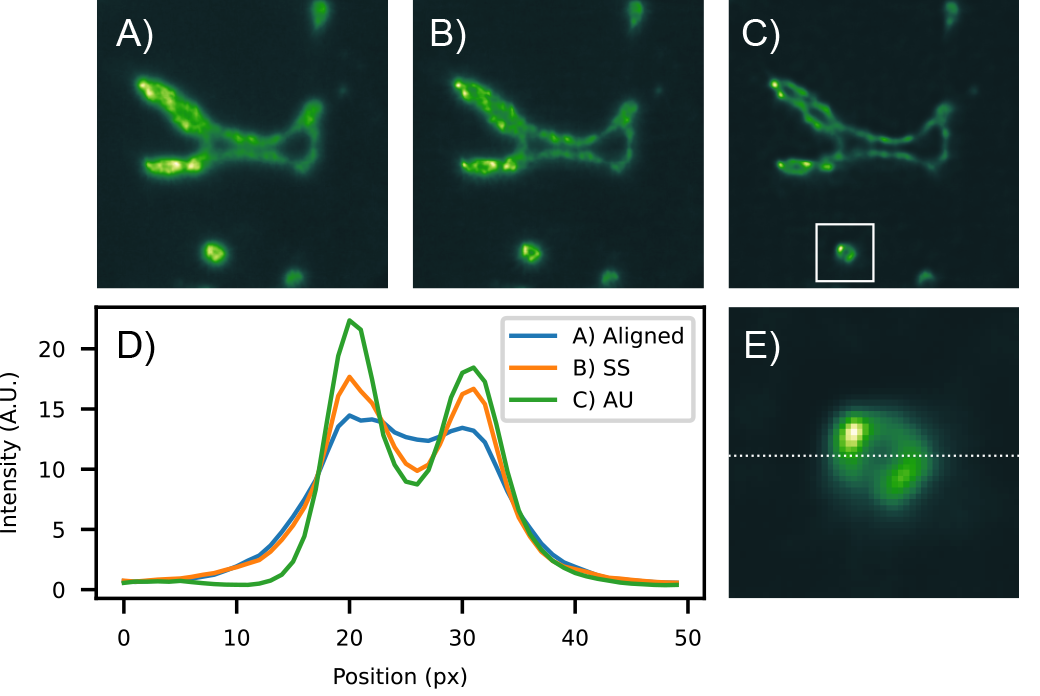}
\caption{Tomographic slice of the cropped volume.
A) Aligned mean (ground truth).
B) Reconstruction using SS.
C) Reconstruction using AU.
D) Profile plot along dashed line in panel E) for the three cases.
E) Detail of the small opening for AU.
}
\label{fig:fig3results}
\end{figure}

We analyze two experimental situations. 
In Fig \ref{fig:fig2maxprojection} and \ref{fig:fig3results}, we report the results obtained on two regions of the specimen. 
The first contains the whole specimen and corresponds to a volume of $512^3$ voxels, with size of $\left(1320\mu m\right)^3$.
The second volume takes a region of interest of $256 \times 256 \times 128$ voxels, with size of $330\mu m \times 330\mu m  \times 165 \mu m$.
Convolutions and correlations are implemented via Fast-Fourier Transform (FFT) spectral decomposition.
The GPU implementation is essential to perform such reconstructions, since the method relies on intensive usage of 3D-FFT.
We implemented the code in Python by using the CuPy library, which provides a flexible CUDA framework for matrix operations.
The problems were tackled using a single nVidia Titan RTX, equipped with $4608$ CUDA cores and $24$ gigabytes of RAM.
For the first volume, each step is accomplished in $0.48$ seconds, while for the second one it needs $0.05$ seconds.
We choose the reference view at angle $\varphi=0\degree$ for the initial guess at $t=0$.
The results obtained for the reconstructions of the whole specimen were rendered in the previous Fig. \ref{fig:fig1schematics}G, where the top-half is the result of SS and the bottom-half is the result of AU.
To compare the different results, we show the maximum intensity projection (MIP) along each spatial coordinate.
The top row of Fig. \ref{fig:fig2maxprojection} shows the ground-truth reconstruction, obtained by averaging the views previously aligned by locating the peak of their cross-correlation.
The second row of Fig. \ref{fig:fig2maxprojection}, instead, shows the results of SS iterations. 
Since the reconstruction is formed from an inherently aligned auto-correlation, the features of the specimen are finely resolved with respect to the ground truth.
Compared to the $\overline{o}_\mu$, the reconstructed $\overline{o}_\rho$ is crisp, with fine features better isolated from an intensity background.
This is due to the sharper PSF $\overline{h}_\text{eff}$ implied by the usage of the auto-correlation.
The third row of Fig. \ref{fig:fig2maxprojection} displays the results obtained with AU, deconvolving $\mathcal{H}$ from the estimated auto-correlation.
The final effect is a deblurring of the reconstruction with respect to the SS.
The validity of the method can be further assessed by examining a tomographic slice taken through the middle of the full-resolution volume.
In Fig. \ref{fig:fig3results}A, we show the ground truth result of the aligned and averaged volume.
We have chosen a detailed region which displays a bifurcated blood vessel, and a smaller circular opening located at the bottom.
Fig. \ref{fig:fig3results}B slices exactly the same plane of the volume $\overline{o}_\rho$ after the inversion of $\overline{\chi}_\mu$ via SS iterations.
If we compare this with the ground-truth, we observe a clear improvement of the reconstruction quality.
Having correctly reinterpreted sub-pixel misalignment and with a neat PSF, the image is rich in details and well contrasted, where the ground truth appears fuzzier.
On the other hand, Fig. \ref{fig:fig3results}C shows the reconstruction of the same volume by using AU.
Here, it is possible to appreciate the deblurring effect that leaves us with a highly-resolved reconstruction.
To assess the qualitative verdict of our analysis, we examine the smaller detail of the vessel located at the bottom of the panel C.
The region of interest is displayed in panel E as a reference in the case of AU reconstruction.
We draw a line profile in the middle of it and we plot the intensities for each of the reconstruction considered in Fig. \ref{fig:fig3results}D.
The ground truth almost confuses the walls of the small blood vessel, whereas SS resolves this detail.
The opening within the blood vessel becomes even more evident in case we use simultaneous deconvolution with AU, given that the deconvolution by the PSF let us resolve sharper details.
A thorough image analysis is presented in the \textcolor{blue}{Supplement Materials} document accompanying this manuscript.

\section*{Discussion}
It is worth stressing that our auto-correlation method goes beyond the deconvolution approach.
We are exploring a new path in alignment-free image formation, studying its advantage in terms of PSF.
With this work, we presented an approach to the problem of shift-invariant reconstructions in volumetric multi-view tomography.
Rather than relying on alignment and fusion pipelines, we proposed a conceptually simpler approach that promotes the reconstruction into the shift-invariant $\mathcal{A}$-space.
We made use of multiple views of the specimen with the sole goal of refining the estimation of the auto-correlation of the object, since we consider it as the ideal quantity for the formation of inherently aligned reconstructions.
By releasing the user from this task, we could direct our attention on better ways to estimate the auto-correlation.
In particular, this may open the path for the corrections of higher-order transformations such as those introduced by inaccuracies of the rotation stage.
For example, two-axes angular tilts are easier to be seen in the shift-invariant space rather than in the object space, since we no longer worry about the object positioning.
Furthermore, we have proven that the solution of the $\mathcal{A}^{-1}$ can be accompanied with deconvolution \cite{ancora2020IEEE}.
Concatenating two inverse problems, in fact, can be hazardous since remaining artifacts from the first inversion may condition the behaviour of the following method.
In volumetric tomography, however, we can always start from a good guess because we have independent object observations which are not too far from the ideal reconstruction.
The initialization with one of these permits us to finalize the reconstruction even if we do not comply with appropriate frequency padding.
In fact, the auto-correlation of a discretely $n$-sampled signal is defined on a translation-space that is $2n-1$ long.
The convergence is guaranteed only if we pad enough the reconstruction volume.
However, to save computer memory, we found that both methods converge if we start with a close guess, i.e. from a single view or the aligned mean.
This is a crucial aspect for the usability of the protocol, because the convergence difficult if we start with a random initialization (in particular with the cropped volume).
However, both SS and AU are quadratic Bayesian methods and typically require many iterations to converge.
With this respect, a new approach to Bayesian deconvolution \cite{guo2020rapid} managed to reduce by two orders of magnitude the iterations needed.
This was done by tuning the forward and backward projection operators, similarly to what we have respectively at the denominator and numerator in our Eq. \ref{eq:anchorUpdate2}.
Reducing the number of iterations is a key aspect that we plan to investigate further in particular for the analysis of big volumes, eventually reducing the GPU processing-time from a few hours to a fraction of it.



\newpage
\section*{Methods}

\subsection*{Correlation/Convolution Theory}
In the following section, we define the quantities used in the manuscript. 
We make use of the bold $\pmb{x}=\left(x,y,z \right)$ to indicate a vector in the direct space $\pmb{x} \in \mathbb{R}^3$. 
Reconstructions are finalized in the direct $\pmb{x}$-space.
Correlations and convolutions are quantities defined by shifts in $\mathbb{R}^3$, which we define with the vector $\pmb{\xi}=\left(\xi, \eta, \zeta \right)$. 
Also $\pmb{\xi} \in \mathbb{R}^3$ and we refer to as the shift-space.

We recall that the Fourier transform $\mathcal{F}$ of a function $h \left( \pmb{x}\right)$ is defined as:
\begin{equation}
    \mathcal{F}\{h\} \equiv \int _{-\infty }^{\infty }h(\pmb{x})\ e^{-i 2\pi \pmb{x} \cdot \pmb{k} }\,d\pmb{x} = {\hat {h}}(\pmb{k} ),
\end{equation}
where $\pmb{k}=\left(k_x,k_y,k_z \right)$ is the wave-vector in frequency space.

\subsubsection*{Cross-correlation}
Conceptually, the cross-correlation is defined as a reference function $o$ multiplied by another function $h$ that shifts by $\pmb{\xi}$ as in:
\begin{equation}
    o\star h = \mathcal{X}\{o;h\} = \int \overline{o\left(\pmb{x} \right)} h\left(\pmb{\xi}+\pmb{x} \right) \, d\pmb{x}.
\end{equation}
The cross-correlation theorem states that $o \star f$ can be written as a product in the Fourier domain:
\begin{equation}
    \mathcal{F}\{o \star h \} = \overline{\mathcal{F}\{o\}} \cdot \mathcal{F}\{h\}.
\end{equation}
For the implementation of our method, we make intensive use of this property and the following Fourier theorems.

\subsubsection*{Auto-correlation}
The auto-correlation is defined as a quantity shifted and multiplied by itself as in:
\begin{equation}
    o\star o = \mathcal{A}\{o\}= \int \overline{o\left(\pmb{x} \right)} o\left(\pmb{\xi}+\pmb{x} \right) \, d\pmb{x} \equiv \chi\left( \pmb{\xi} \right).
\end{equation}
The Wiener-Khinchin (power-spectrum) theorem states:
\begin{equation}
    \mathcal{F}\{o\star o \} = \| \mathcal{F}\{o\} \|^2.
\end{equation}
The auto-correlation of any real signal is an even function, thus $\chi\left( \pmb{\xi} \right)=\chi\left(- \pmb{\xi} \right)$, and it is always centered around its maximum located at $\pmb{\xi}=\left(0,0,0 \right) \equiv \pmb{0}$.

\subsubsection*{Convolution}
The convolution is defined by shifting and multiplying two functions $o$ and $h$ as in:
\begin{equation}
    o*h = \mathcal{C}\{o;h\} = \int o\left(\pmb{x} \right) h\left(\pmb{\xi}-\pmb{x} \right) \, d\pmb{x}.
\end{equation}
The convolution theorem states that:
\begin{equation}
    \mathcal{F}\{o*h\} = \mathcal{F}\{o\} \cdot \mathcal{F}\{h\}.
\end{equation}

\subsection*{Image pre-processing}
Each raw dataset was subtracted with a corresponding average background value, rotated to the same angular orientation of the first dataset acquired at $\varphi=0$.
The PSF of the system was assumed to be Gaussian, elongated along the direction of scanning. 
For each of these stacks, we computed the corresponding auto-correlation sequence.
We took the absolute value of the average auto-correlation to avoid any presence of unwanted negative values.
These are determined by the background-subtraction and eventually by rounding errors due to FFT computation.

\subsection*{Multi-view registration and fusion}
As ground truth comparison, we aligned the views against each other by finding the location of the maximum of the cross-correlation between $\mathcal{X}\{o_\mu^i;o_\mu^j\}$, for $i \neq j$.
We defined the displacement vector $\pmb{m}_i$ with respect the central coordinate $\pmb{\xi}=\pmb{0}$.
We kept $\varphi_{i=0}=0\degree$ as reference and we translated each $\varphi_i$ by the vector $-\pmb{m}_i$ defined in this way.
Then we computed the average of the registered stacks to form $\overline{o}_\mu$. 

\subsection*{Rearranging the auto-correlation}
The measurement $o_\mu$:
\begin{equation}
    o_\mu = o * h + \varepsilon.
\end{equation}
We neglect the additive noise by assuming $\varepsilon=0$.
Using the commutation properties of the convolution and correlation, we have that:
\begin{align}
    \chi_\mu &\equiv \mathcal{A}\{o_\mu\} = o_\mu \star o_\mu = \left(o*h \right) \star \left(o*h \right) \\ 
    &= \left( o\star o \right)  * \left(h\star h \right) = o * \left(o \star \left(h\star h \right) \right) \label{eq:blurredACorr2}\\
    &= \chi * \mathcal{H} = o * \mathcal{K}. 
\end{align}
Here we have called $\chi=o\star o$, $\mathcal{H}=h\star h$ and $\mathcal{K}=o \star \mathcal{H}$.
In the main text, we use only the last two equations.

\subsection*{Experimental details}
For the test performed in our letter, we use a cleared mouse popliteal lymph node having the vasculature stained with the Alexa Fluor 488 dye.
The sample was embedded in agarose, then cleared and imaged in Benzyl-Alcohol Benzyl-Benzoate (BABB).
The fluorescence is excited with a light-sheet perpendicular to the camera detection at  $\lambda_{exc} = 488nm$, imaged onto the sample with an 2.5x/0.07 N PLAN (air) objective lens.
With a band-pass filter at $525/50nm$, we imaged the emitted fluorescence using a 5x/0.12 N Plan EPI (air) objective lens.
The sample was scanned through the light sheet along a z-axis, perpendicularly to the camera detection in steps of $4.985\mu m$.
The specimen was provided by Prof. J. Stein at the University of Bern and imaged by Jim Swoger at the Center for Genomic Regulation (CRG), Barcelona.

\newpage

\section*{Fundings}
H2020 Marie Skłodowska-Curie Actions (HI-PHRET project, 799230); H2020 Laserlab Europe V (871124).

\section*{Acknowledgment}
Sample provided by Prof. J. Stein at the University of Bern and imaged by Jim Swoger at the Center for Genomic Regulation (CRG), Barcelona.
The authors further thank Dr. Gianmaria Calisesi for inspiring discussions.

\section*{Author contributions}
You must include a statement that specifies the
individual contributions of each co-author. For example: "A.P.M.
‘contributed’ Y and Z; B.T.R. ‘contributed’ Y,” etc. See our authorship
policies for more details.

\section*{Competing interests}
The authors declare no competing interests.

\section*{Materials \& Correspondence}
Correspondence to Daniele Ancora.

\newpage

\bibliography{bibliography.bib}

\begin{thebibliography}{10}
\urlstyle{rm}
\expandafter\ifx\csname url\endcsname\relax
  \def\url#1{\texttt{#1}}\fi
\expandafter\ifx\csname urlprefix\endcsname\relax\def\urlprefix{URL }\fi
\expandafter\ifx\csname doiprefix\endcsname\relax\def\doiprefix{DOI: }\fi
\providecommand{\bibinfo}[2]{#2}
\providecommand{\eprint}[2][]{\url{#2}}

\bibitem{sun2019summarizing}
\bibinfo{author}{Sun, Y.}, \bibinfo{author}{Agostini, N.~B.},
  \bibinfo{author}{Dong, S.} \& \bibinfo{author}{Kaeli, D.}
\newblock \bibinfo{journal}{\bibinfo{title}{Summarizing cpu and gpu design
  trends with product data}}.
\newblock {\emph{\JournalTitle{arXiv preprint arXiv:1911.11313}}}
  (\bibinfo{year}{2019}).

\bibitem{leiserson2020there}
\bibinfo{author}{Leiserson, C.~E.} \emph{et~al.}
\newblock \bibinfo{journal}{\bibinfo{title}{There’s plenty of room at the
  top: What will drive computer performance after moore’s law?}}
\newblock {\emph{\JournalTitle{Science}}} \textbf{\bibinfo{volume}{368}}
  (\bibinfo{year}{2020}).

\bibitem{despres2017review}
\bibinfo{author}{Despres, P.} \& \bibinfo{author}{Jia, X.}
\newblock \bibinfo{journal}{\bibinfo{title}{A review of gpu-based medical image
  reconstruction}}.
\newblock {\emph{\JournalTitle{Physica Medica}}} \textbf{\bibinfo{volume}{42}},
  \bibinfo{pages}{76--92} (\bibinfo{year}{2017}).

\bibitem{sharpe2002optical}
\bibinfo{author}{Sharpe, J.} \emph{et~al.}
\newblock \bibinfo{journal}{\bibinfo{title}{Optical projection tomography as a
  tool for 3d microscopy and gene expression studies}}.
\newblock {\emph{\JournalTitle{Science}}} \textbf{\bibinfo{volume}{296}},
  \bibinfo{pages}{541--545} (\bibinfo{year}{2002}).

\bibitem{verveer2007high}
\bibinfo{author}{Verveer, P.~J.} \emph{et~al.}
\newblock \bibinfo{journal}{\bibinfo{title}{High-resolution three-dimensional
  imaging of large specimens with light sheet--based microscopy}}.
\newblock {\emph{\JournalTitle{Nature methods}}} \textbf{\bibinfo{volume}{4}},
  \bibinfo{pages}{311--313} (\bibinfo{year}{2007}).

\bibitem{krzic2012multiview}
\bibinfo{author}{Krzic, U.}, \bibinfo{author}{Gunther, S.},
  \bibinfo{author}{Saunders, T.~E.}, \bibinfo{author}{Streichan, S.~J.} \&
  \bibinfo{author}{Hufnagel, L.}
\newblock \bibinfo{journal}{\bibinfo{title}{Multiview light-sheet microscope
  for rapid in toto imaging}}.
\newblock {\emph{\JournalTitle{Nature methods}}} \textbf{\bibinfo{volume}{9}},
  \bibinfo{pages}{730--733} (\bibinfo{year}{2012}).

\bibitem{weber2012omnidirectional}
\bibinfo{author}{Weber, M.} \& \bibinfo{author}{Huisken, J.}
\newblock \bibinfo{journal}{\bibinfo{title}{Omnidirectional microscopy}}.
\newblock {\emph{\JournalTitle{Nature methods}}} \textbf{\bibinfo{volume}{9}},
  \bibinfo{pages}{656} (\bibinfo{year}{2012}).

\bibitem{swoger2007multi}
\bibinfo{author}{Swoger, J.}, \bibinfo{author}{Verveer, P.},
  \bibinfo{author}{Greger, K.}, \bibinfo{author}{Huisken, J.} \&
  \bibinfo{author}{Stelzer, E.~H.}
\newblock \bibinfo{journal}{\bibinfo{title}{Multi-view image fusion improves
  resolution in three-dimensional microscopy}}.
\newblock {\emph{\JournalTitle{Optics express}}} \textbf{\bibinfo{volume}{15}},
  \bibinfo{pages}{8029--8042} (\bibinfo{year}{2007}).

\bibitem{preibisch2014efficient}
\bibinfo{author}{Preibisch, S.} \emph{et~al.}
\newblock \bibinfo{journal}{\bibinfo{title}{Efficient bayesian-based multiview
  deconvolution}}.
\newblock {\emph{\JournalTitle{Nature methods}}} \textbf{\bibinfo{volume}{11}},
  \bibinfo{pages}{645} (\bibinfo{year}{2014}).

\bibitem{ancora2017phase}
\bibinfo{author}{Ancora, D.} \emph{et~al.}
\newblock \bibinfo{journal}{\bibinfo{title}{Phase-retrieved tomography enables
  mesoscopic imaging of opaque tumor spheroids}}.
\newblock {\emph{\JournalTitle{Scientific reports}}}
  \textbf{\bibinfo{volume}{7}}, \bibinfo{pages}{11854} (\bibinfo{year}{2017}).

\bibitem{ancora2018optical}
\bibinfo{author}{Ancora, D.} \emph{et~al.}
\newblock \bibinfo{journal}{\bibinfo{title}{Optical projection tomography via
  phase retrieval algorithms}}.
\newblock {\emph{\JournalTitle{Methods}}} \textbf{\bibinfo{volume}{136}},
  \bibinfo{pages}{81--89} (\bibinfo{year}{2018}).

\bibitem{shechtman2015phase}
\bibinfo{author}{Shechtman, Y.} \emph{et~al.}
\newblock \bibinfo{journal}{\bibinfo{title}{Phase retrieval with application to
  optical imaging: a contemporary overview}}.
\newblock {\emph{\JournalTitle{IEEE signal processing magazine}}}
  \textbf{\bibinfo{volume}{32}}, \bibinfo{pages}{87--109}
  (\bibinfo{year}{2015}).

\bibitem{schindelin2012fiji}
\bibinfo{author}{Schindelin, J.} \emph{et~al.}
\newblock \bibinfo{journal}{\bibinfo{title}{Fiji: an open-source platform for
  biological-image analysis}}.
\newblock {\emph{\JournalTitle{Nature methods}}} \textbf{\bibinfo{volume}{9}},
  \bibinfo{pages}{676--682} (\bibinfo{year}{2012}).

\bibitem{schulz1992image}
\bibinfo{author}{Schulz, T.~J.} \& \bibinfo{author}{Snyder, D.~L.}
\newblock \bibinfo{journal}{\bibinfo{title}{Image recovery from correlations}}.
\newblock {\emph{\JournalTitle{JOSA A}}} \textbf{\bibinfo{volume}{9}},
  \bibinfo{pages}{1266--1272} (\bibinfo{year}{1992}).

\bibitem{ancora2020IEEE}
\bibinfo{author}{Ancora, D.} \& \bibinfo{author}{Bassi, A.}
\newblock \bibinfo{journal}{\bibinfo{title}{Deconvolved image restoration from
  auto-correlations}}.
\newblock {\emph{\JournalTitle{IEEE Transactions on Image Processing}}}
  \textbf{\bibinfo{volume}{30}}, \bibinfo{pages}{1332--1341}
  (\bibinfo{year}{2020}).

\bibitem{guizar2008efficient}
\bibinfo{author}{Guizar-Sicairos, M.}, \bibinfo{author}{Thurman, S.~T.} \&
  \bibinfo{author}{Fienup, J.~R.}
\newblock \bibinfo{journal}{\bibinfo{title}{Efficient subpixel image
  registration algorithms}}.
\newblock {\emph{\JournalTitle{Optics letters}}} \textbf{\bibinfo{volume}{33}},
  \bibinfo{pages}{156--158} (\bibinfo{year}{2008}).

\bibitem{guo2020rapid}
\bibinfo{author}{Guo, M.} \emph{et~al.}
\newblock \bibinfo{journal}{\bibinfo{title}{Rapid image deconvolution and
  multiview fusion for optical microscopy}}.
\newblock {\emph{\JournalTitle{Nature Biotechnology}}}
  \textbf{\bibinfo{volume}{38}}, \bibinfo{pages}{1337--1346}
  (\bibinfo{year}{2020}).

\end{thebibliography}

\end{document}